\documentclass[conference]{IEEEtran}
\newif\ifblinded\blindedfalse
\usepackage[utf8]{inputenc}

\usepackage{cite}
\usepackage{amsmath,amssymb,amsfonts}
\usepackage{graphicx}
\usepackage{textcomp}

\usepackage{booktabs}
\usepackage{bytefield}
\usepackage{color,colortbl}
\usepackage{xspace}
\usepackage{tikz}
\ifcsname basic.pgf\endcsname
  \expandafter 
\fi
\expandafter\gdef\csname basic.pgf\endcsname{loaded}

\usetikzlibrary{calc}
\usetikzlibrary{shapes.geometric}
\usetikzlibrary{circuits.logic.US}

\newcommand{\rsec}[1]{Sec.\,\ref{#1}}
\newcommand{\rfig}[1]{Fig.\,\ref{#1}}
\newcommand{\rtab}[1]{Tab.\,\ref{#1}}

\renewcommand\bar\overline

\title{%
  Using DSP Slices as\\Content-Addressable Update Queues
}
\author{%
\ifblinded\else
  \IEEEauthorblockN{%
    Thomas B. Preu{\ss}er\IEEEauthorrefmark{1},
    Monica Chiosa\IEEEauthorrefmark{1},
    Alexander Weiss\IEEEauthorrefmark{2},
    Gustavo Alonso\IEEEauthorrefmark{1}
  }
  \IEEEauthorblockA{%
    \IEEEauthorblockA{\IEEEauthorrefmark{1}%
        Systems Group, Department of Computer Science, ETH Zürich, Switzerland
    }
    \IEEEauthorblockA{\IEEEauthorrefmark{2}%
        Accemic Technologies, Kiefersfelden, Germany
    }
  }
\fi
}

\begin{document}
\maketitle

\begin{abstract}
Content-Addressable Memory (CAM) is a powerful abstraction for building
memory caches, routing tables and hazard detection logic. Without a native
CAM structure available on FPGA devices, their functionality must be emulated
using the structural primitives at hand. Such an emulation causes
significant overhead in the consumption of the underlying resources,
typically general-purpose fabric and on-chip block RAM (BRAM). This often
motivates mitigating trade-offs, such as the reduction of the
associativity of memory caches. This paper describes a technique to implement
the hazard resolution in a memory update queue that hides the off-chip memory
readout latency of read-modify-write cycles while guaranteeing the delivery
of the full memory bandwidth. The innovative use of DSP slices allows them
to assume and combine the functions of (a) the tag and data storage,
(b) the tag matching, and (c) the data update in this key-value storage scenario.
The proposed approach provides designers with extra flexibility by adding this
resource type as another option to implement CAM.
\end{abstract}


\section{Introduction}
Content-Addressable Memory (CAM) provides
valuable functionality at the core of many practical use cases. CAMs
allow to associate some data with a set of keys, aka. tags, coming from a
considerably larger key space. They are found (a) at the heart of
memory caches that buffer the contents of a small subset of memory
addresses, (b) in the implementations of network routing tables which
typically support ternary wildcard matching, and (c) in the
hazard detection logic of command queues of long-latency off-chip
memory controllers. Depending on the use case, the associated value
may be (a) a data vector such as a cache line, (b) a scalar value as for
identifying a forwarding interface, or (c) a simple flag as for
indicating that a memory location is currently manipulated.

The implementation of CAM structures on FPGAs is challenging. It
requires the efficient integration of both storage capabilities and
vastly parallel key matching logic. While modern FPGAs comprise sizable
designated on-chip memory in the form of block RAM (BRAM) or UltraRAM
(URAM), these RAM abstractions encounter a bottleneck in the number of
available memory ports and cannot facilitate the parallel comparison
of their contents against a lookup key.

Another hardened special-purpose structure embedded into the fabric of
virtually all modern FPGAs is the DSP block \cite{ug479,ug579}. It is
best known for comprising a hard-wired word multiplier that can be
used for scalar multiplication or put on an accumulation chain together
with other DSP blocks to compute dot products.

Applications with little or no need for multipliers often let the DSP
blocks of the device sit idle. However, even disregarding the
multiplier, the DSP blocks provide interesting infrastructure involving
wide datapaths, both internal to a block and cascaded to its
neighbors, with an abundance of pipeline registers. Furthermore, simple
arithmetic, such as addition and subtraction, bitwise logic
operations, as well as word comparison are available independently of
the multiplier. These capabilities are an interesting foundation for
implementing an integrated key storage and matching logic and render DSP slices
an attractive alternative to claiming vast amounts of fabric LUTs,
registers, and general-purpose routing. Note that DSP blocks can be
operated in a mode (\texttt{USE\_MULT="NONE"}) that deactivates the
multiplier so as to eliminate the power overhead of the unused multiplier.

This paper reviews different techniques used to leverage various
FPGA resource types for implementing CAMs. It then makes the case
for using DSP slices as another point in the design space offering
interesting resource trade-offs. In the rest of this paper,
\rsec{secRelated} gives an overview of the state of the art before
\rsec{secUseCase} details the constraints and functional
requirements of a use case serving as a running example.  Our
DSP-based solution is presented in \rsec{secImplementation} and
evaluated in \rsec{secEvaluation}. Finally, \rsec{secConclusion}
concludes the paper.

\section{Related Work}\label{secRelated}
FPGAs lack hardened CAM structures so that CAM functionality must be
emulated. Since employing LUTs and registers is extremely hungry in terms of
resources, RAM-based emulation has become a popular
technique to implement CAMs. Using the key for an address-based lookup, however, quickly
encounters scalability bounds. Zerbini and Finochietto
\cite{zerbini:2012} demonstrate that slicing the key into individual
subwords allows to partition the lookup using separate but
significantly smaller memories. Indeed, slicing the key into
individual bits would match the structure of a native CAM that
determines the match from an \texttt{AND} tree combining the match
results of all key bits. This approach has been elaborated by Jiang
\cite{jiang:2013} who formalizes the approach and adds an extensive
experimental evaluation. Both of them consider a networking background
and assume an implementation as TCAM that also supports ternary
wildcard matches. The TCAM-nature of their emulators renders
updates more expensive, requiring $O\left(2^w\right)$ RAM writes with
$w$ being the bit length of the longest key slice
\cite{jiang:2013}. This update time can, in fact, be reduced to $O(1)$
when wildcard matching is not needed. In any case, the length of the
key slices induces a memory space overhead factor of $\frac{2^w}{w}$
over a native CAM \cite{jiang:2013}. For the typical fabric memory
resources, this implies a $\frac{32}{5}\times=6.4\times$ overhead for
LUTRAM and a $\frac{512}{9}\times=56.9\times$ overhead for BRAM.

Large associative mappings require to leverage FPGA-attached off-chip
memory, typically DRAM. Use cases include FPGA-accelerated
network routing \cite{bando:2009}, particularly for IPv6, or key-value
stores \cite{blott:2013}. While Bando et\,al. \cite{bando:2009} and
Blott et\,al. \cite{blott:2013} use hashing to randomize storage,
trie-based IP routing implementations have also been reported
\cite{song:2005,jiang:2008}. The practical challenges in these
approaches are the hash collision management and the accumulating
pointer chasing latency, respectively. Such large, or often even
canonical mappings, with a rather high tolerance for lookup latency
are not targeted by the solution offered in this paper.

Another use case frequently associated with CAMs is memory caching. As
caching is a performance-enhancing technique that tolerates cache misses,
its implementation can include more compromises and trade-offs. The critical tuning
parameter is, in fact, associativity. Direct-mapped or set-associative
implementations with a low associativity of two or four effectively
reduces the tag lookup parallelism, thereby enabling a mapping to on-chip
RAM resources. The focus of FPGA optimization efforts then moves to
device-compatible replacement strategies \cite{xue:2014,titinchi:2019}
and non-blocking implementations that are able to handle parallel outstanding
misses \cite{dundas:1997,asiatici:2019}.

Hazard detection and prevention is used to guarantee the atomicity of
spatially or temporally interleaved data write and read accesses. As
long as blocking is tolerable, only the keys under pending updates
must be identified safely. Since false positives are functionally
tolerable in this set membership formulation, Bloom filters
\cite{bloom:1970}, in particular counting Bloom filters
\cite{fan:1998} that support set removal, are a very efficient
implementation option. If updates are rare, even more primitive data
structures such as sequentially iterated lists may offer a feasible
solution as practiced in the key-value store Caribou
\cite{istvan:2018}.

\section{Use Case \& Challenge}\label{secUseCase}
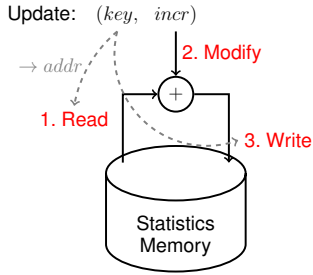
\begin{figure}
  \centerline{\resizebox{.5\linewidth}{!}{% Author: Thomas B. Preußer
\usetikzlibrary{calc}
\usetikzlibrary{shapes.geometric}

\begin{tikzpicture}[scale=.9,line width=1.2,font=\large\sffamily]
\node(mem) at(0,0) [cylinder,draw,aspect=.5,rotate=90,minimum height=27mm,minimum width=30mm]{\rotatebox{270}{\raisebox{-8pt}[0pt][0pt]{\begin{tabular}{c}Statistics\\Memory\end{tabular}}}};
\node(add)[circle,draw]at(0,3.2){$+$};
\draw[->] ($(mem.north east)+(.3,0)$)|-(add);
\draw[<-] ($(mem.south east)+(-.3,0)$)|-(add);
\draw[<-](add)--+(0,1.5) node(incr)[above]{$incr)$} node[midway,right,red]{2. Modify};
\node(addr) at($(incr)-(1.4,0)$) {$(key,$};
\node at($(addr)-(1.8,0)$) {Update:};
\draw[dashed,->,gray](addr.south) arc (130:170:3) node[below,red]{1. Read} node[midway,left]{$\to addr$};
\draw[dashed,->,gray](addr.south) arc (166:290:2.2) node[right,red]{3. Write};
\end{tikzpicture}}}
  \caption{Off-Chip Statistics Collection}
  \label{figUseCase}
\end{figure}
Our reference use case is the online collection of event stream
statistics. As shown in \rfig{figUseCase}, events serve as the key
that is mapped to an associated value, e.g., their occurrence count. A
statistics update requires (a) the current value to be read, (b) the
update to be applied and (c) the updated value to be written
back. While off-chip memory is able to support a space of
multi-million keys, the practical implementation of the required
read-modify-write cycle must cope with the rather high read latency of
this memory. In particular, updates may not be allowed to cause the
loss of a previous one by issuing a stale read interleaved with an
ongoing, not yet completed update cycle. Indeed, Bloom filters may be
used to enforce the sequentialization of updates to the same
key. However, this approach will cripple the utilized memory bandwidth
and expose the complete memory read latency as event input period
under key locality.

The key measure to maintain the event throughput is to conflate
secondary updates to a reoccurring key with pending ones. Instead of issuing a read
for starting a fresh memory update cycle, the outstanding write back will be
manipulated to perform a cumulative update. The required state can be
captured by an on-chip mapping associating keys within an update cycle with
their scheduled transition function, such as an increment.
This mapping must be able to support a subset of the key space
whose size, at least, reflects the average read latency of a few
100\,ns exposed by the underlying memory controller. Over-provisioning
the mapping size helps to hide individual tail
latencies as induced by bus turnarounds or memory refreshes. At
practical application clock frequencies between 200\,MHz and
400\,MHz, this amounts to maintaining a few hundred outstanding key-update
mappings with fully associative matching capabilities.

The described use case comes with a few distinct properties that can
be exploited by an implementation:
\begin{itemize}
\item
  The insertion order equals the deletion order as (a) an insertion
  reflects a read request issued for a key without an outstanding
  update and (b) its deletion represents the write back upon the
  reception of the corresponding reply.
\item
  Write backs may be delayed beyond the reception of the current
  memory value as long as the update conflation is continued and the
  write back is performed eventually before using the statistics
  accumulated in memory.
\end{itemize}
The first property allows the mapping to be structured as a queue. The
second one permits a static depth, which may be used for structural
simplification.

\section{Implementation Concept}\label{secImplementation}
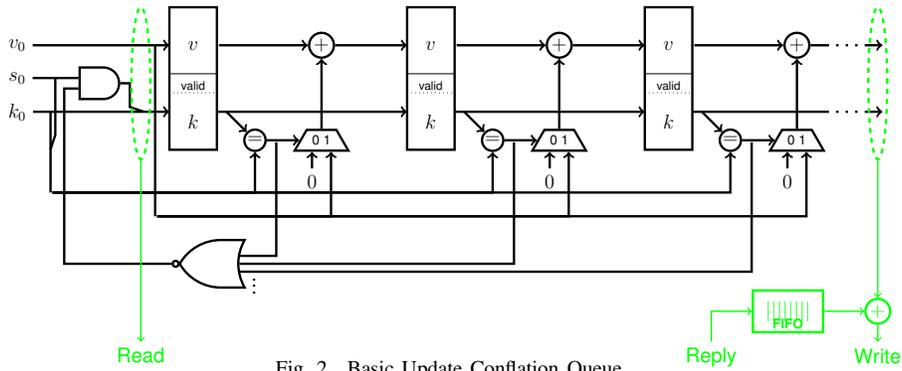
\begin{figure*}
  \centerline{\resizebox{.68\linewidth}{!}{\usetikzlibrary{calc}
\usetikzlibrary{circuits.logic.US}
\usetikzlibrary{shapes.geometric}

\begin{tikzpicture}[circuit logic US,scale=1,line width=1.4,font=\large\sffamily]

% Inputs
\node(and)  at(-1.5,1.4) [and gate, minimum height=6mm,rotate=0] {};

\node(nval_in) at(-3.2,2.2) {$v_0$}; \draw[->](nval_in) -- (nval_in-|0,0);
\draw (nval_in-|-.3,0) -- +(0,-3.6) node[inner sep=0](nval_src) {};

\node(nstb_in) at(nval_in|-and.input 1) {$s_0$}; \draw(nstb_in) -- (and.input 1);

\node(nkey_in) at(nval_in|-0,0.8) {$k_0$}; \draw[->](nkey_in) -- (nkey_in-|0,0);
\draw ($(nkey_in)+(.7,0)$) -- +(0,-1.7) node[inner sep=0](nkey_src) {};
\draw ($(nstb_in)+(.8,0)$) -- ++(0,-1.2) -- +(-.1,-.3);
\draw (and.output) -| ++(.2,-.5) -- +(.3,-.1);

\foreach \x in {0,1,2} {

  % Register
  \draw (5*\x,0) rectangle +(1,3);
  \draw[thin,dotted] (5*\x,1.2) -- +(1,0);
  \draw[thin] (5*\x,1.6) -- +(1,0);
  \node at(5*\x+.5,.6) {$k$};
  \node at(5*\x+.5,1.36) {\scriptsize valid};
  \node at(5*\x+.5,2.2) {$v$};

  \draw[->] (5*\x+1,.8) node[inner sep=0] (nks) {} -- +(4,0);
  \node[inner sep=0] (nvs) at (5*\x+1,2.2) {};

  % Datapath
  \node[draw,circle,inner sep=1] (ncmp) at(5*\x+1.8,.2){$=$};
  \draw[->] ($(nks)+(.2,0)$) -- (ncmp);
  \draw[->] (nkey_src) -| (ncmp);
  \node[draw,trapezium] (nsel) at($(ncmp)+(1.4,0)$) {\footnotesize 0 1};
  \draw[->] (ncmp) -- (nsel);
  \draw[<-] ($(nsel.south)+(-.2,0)$) -- +(0,-.35) node[below] {$0$};
  \draw[->] (nval_src)-|($(nsel.south)+(.2,0)$);
  \node[inner sep=0] (nmatch\x) at ($(ncmp.east)+(.2,0)$) {};

  \node[draw,circle,inner sep=1] (nadd) at(nvs-|nsel){$+$};
  \draw[->] (nsel) -- (nadd);
  \draw[->] (nvs) -- (nadd);
  \draw[->] (nadd) -- ($(nvs)+(4,0)$);
}

\node(nor)  at(1,-2.4) [nor gate, minimum height=10mm,rotate=180] {};
\draw(nmatch0) |- (nor.input 2);
\draw(nmatch1) |- (nor);
\draw(nmatch2) |- (nor.input 1);
\node at($(nor.input 1)+(.3,-.2)$) {$\vdots$};

\draw (nor.output) -| ($(and.input 2)-(.3,0)$) -- (and.input 2);

\draw[fill,white] (14.5,0) rectangle ++(.1,3);
\draw[fill,white] (14.3,0) rectangle ++(.1,3);
\draw[fill,white] (14.1,0) rectangle ++(.1,3);
\draw[fill,white] (13.9,0) rectangle ++(.1,3);

\draw[color=green,dashed] (-.6,1.4) ellipse (.2 and 1.6);
\draw[color=green,dashed] (14.9,1.4) ellipse (.2 and 1.6);

% Write Back
\node(nfifo)[right,draw,green,minimum height=25,minimum width=42] at(0,-3.4-|nmatch2) {$| | | | | | | |$};
\node[green,below] at(nfifo) {\textbf{\footnotesize FIFO}};
\node(ntotal)[draw,green,circle,inner sep=1] at(nfifo-|14.9,0) {$+$};
\draw[->,thick,green] (nfifo) -- (ntotal);
\draw[->,thick,green] (14.9,-.2) -- (ntotal);
\draw[->,thick,green] (ntotal) -- +(0,-.64) node(nwr) {};
\node[below,green] at(nwr) {Write};

% Read & Reply
\draw[->,thick,green] (-.6,-.2) -- (-.6,0|-nwr) node(nrd)[below] {Read};
\draw[<-,thick,green] (nfifo)-|(11.4,0|-nwr)node[green,below] {Reply};

\end{tikzpicture}}}
  \vspace*{-6mm}
  \caption{Basic Update Conflation Queue}
  \label{figBasicQueue}
\end{figure*}
The basic idea for the update conflation queue is illustrated in
\rfig{figBasicQueue}. A pipeline whose length is matched to the read
latency of the off-chip memory carries tuples comprising keys ($k$) and
the increment values ($v$). The pipeline never stalls unless forced so by
backpressure from the memory controller. It may transport empty slots,
which are identified by a deasserted \emph{valid} flag, which is prepended to each
key and included in key matching. Empty slots arise (a) from cycles without
an update input (strobe $s_0=0$) or (b) from updates that have been conflated
with a pending one. An update conflation is triggered \emph{locally}: when
the parallel matching of the new key $k_0$ with all pending updates
encounters a match, the outstanding increment value will subsume the new
update $v_0$. If any match has been determined, this information will be fed
back through the NOR gate. Only unmatched keys (a) effect a memory read
for initiating a read-modify-write cycle and (b) create a valid key-value
pair in the conflation queue. Once this pair has propagated through the
entire queue, subsuming all updates to the same key during propagation, the
memory reply will be available. The overall increment can be performed
and written back to external memory.

While the conflation of updates for the same key may reduce the required
memory bandwidth, an update stream of distinct keys without gaps will still
emit two memory operations, a read and a write, for each clock cycle. As
memories are optimized for throughput and expose multi-lane interfaces to a
user design, this is not an obstacle per se. However, memory operations
should be grouped by kind, i.e. read or write, to reduce bus turnaround
overhead. It is critical in any reordering of commands to achieve such a
grouping that a read never overtakes a write to the same address. A
scheduling that strictly does not allow any reads to pass writes is a
simple approach to satisfy this demand. Note that these
considerations apply to any implementation regardless of how it chooses to
manage read-modify-write cycles with pending memory reads.

The operations needed for the update conflation for each stage of
\rfig{figBasicQueue} are all available in a DSP slice. However, its completely
combinatorial matching is impractical. Firstly, it would result in prohibitively
long critical path delays and, hence, an unattractive clock frequency. More
fundamentally, the computation must be re-arranged to match the structural
composition of a DSP slice. Note particularly that the comparator result is
registered along with the accumulated output \cite{ug479,ug579}.
This register delay requires the addition merging two matching updates
to occur one pipeline stage later.

\begin{figure}
  \resizebox{\linewidth}{!}{% Author: Thomas B. Preußer
\usetikzlibrary{calc}
\usetikzlibrary{circuits.logic.US}
\usetikzlibrary{shapes.geometric}

\begin{tikzpicture}[circuit logic US,scale=.9,line width=1.2]

% Input Lines
\draw(-1.8,-1.5)node(k0)[below]{$k_{-3})$} --+(0,12);
\draw(-2.8,-1.5)node(v0)[below]{$(v_{-3},$}--+(0,12);

% DSP Blocks: Frames
\foreach \i in {0,1} {
\pgfmathparse{5.8*\i+2.3*int(div(\i,2))}
\begin{scope}[shift={(0,\pgfmathresult)}]
\draw[line width=2.4pt,gray,rounded corners](-1,-1.2) rectangle +(11.6,5.8) node[below left,xshift=-4mm] {\textbf{\large DSP}};
\end{scope}
}

% DSP Blocks: Interior
\newcommand\vc[2]{\left(\begin{array}{@{}c@{}}#1\\#2\end{array}\right)}
\foreach \i in {1,0} {
\pgfmathparse{5.8*\i+2.3*int(div(\i,2))}
\begin{scope}[shift={(0,\pgfmathresult)}]

% Input Registers
\node[draw,line width=2,rectangle,minimum width=10mm,minimum height=14mm](ab1) at (0.0,3) {A:B};
\node[draw,line width=2,rectangle,minimum width=10mm,minimum height=14mm](ab2) at (2.8,3) {A:B};
\draw[<-](ab1)--(ab1-|v0);

% MUX
\node[draw,line width=2,rotate=270,trapezium,trapezium angle=60,minimum width=18mm](kill\i) at (5.6,2.5) {};
\node[right] at(kill\i.south west) {$1$};
\node[right] at(kill\i.south east) {$0$};
\draw[<-](kill\i.south east)--+(-.5,0) node[left] {$\ifnum\i=0\relax\vc{\cdot}{\overline{K}}\else0\fi$};

% Compute
\node[draw,line width=2,circle,minimum size=6mm](sum\i) at (7,2.5) {$+$};
\node[draw,line width=2,rectangle,minimum width=8mm,minimum height=14mm](p\i) at (9.5,2.5) {P};
\draw[->](ab1)--(ab2)					node[midway,above] {$\vc{v_{-2}}{\ifnum\i=0\relax k_{-2}\else0\fi}$};
\draw[->](ab2)--(kill\i.south west)	node[midway,above] {$\vc{v_{-1}}{\ifnum\i=0\relax k_{-1}\else0\fi}$};
\draw[->](kill\i)--(sum\i);
\draw[->](sum\i)--(p\i);

\ifnum\i=0\draw[<-](sum\i)-- +(0,-1) node[below]{$0$};\else\draw[<-](sum\i)-- ++(0,-3)-|++(3.3,-0.7)node(ss\i){};\fi
\ifnum\i>0\draw[<-](kill\i)--++(0,-3.4)-|++(5.6,-0.3)node(kk\i){};\fi

\ifnum\i=3\relax
\draw[->](p\i)--+(3,0) node[above]{$\vc{v_n}{k_n}$};
\else
\draw[->](p\i) -| +(0.8,2.1) node(pp\i){};
% Compare
\node[draw,line width=2,rectangle,minimum width=8mm,minimum height=14mm](c) at (0,0) {C};
\draw[<-](c)--(c-|k0);
\node[draw,line width=2,circle,minimum size=6mm](cmp) at (8,1) {$=$};
\node[draw,line width=2,rectangle,minimum width=8mm,minimum height=8mm](m\i) at (9.5,1) {Q};
\draw[->](sum\i)-|(cmp);
\draw[->](c)-|(cmp) node[very near start,above] {$\vc{\cdot}{k_{-2}}$};
\draw[->](cmp)--(m\i);
\draw[->](m\i) -| +(1.7,3.6) node(mm\i){};

% Outputs and Labels
\draw[->](m\i)--+(2.8,0) node(q\i){};\node[above]at(q\i){$k_{-1}=k_{\ifnum\i<2\relax\i\else n-1\fi}$};
\node at(q\i|-p\i) {$\vc{v_{\ifnum\i<2\relax\i\else n-1\fi}}{k_{\ifnum\i<2\relax\i\else n-1\fi}}$};
\fi
\end{scope}
}
\node at(5,11) {\Huge$\vdots$};
\node(or) at(13.4,0)[nor gate,minimum height=10mm,rotate=-90]{};
\draw(q0) -| (or.input 2);
\draw(q1) -| (or);
\node[above] at(or.north west){\textbf{\dots}};
\draw[->](or.output)--+(0,-1)-|(kill0) node(nm)[midway,left]{$\lnot\exists i\in\left[0,n\right).k_{-1}=k_i$};

\node[right,green] at (ab1-|v0) {\textbf{//}};
\node[green] at (nm-|sum0) {\textbf{//}};
\end{tikzpicture}}
  \caption{DSP Mapping of Conflation Queue}
  \label{figDSPMapping}
\end{figure}
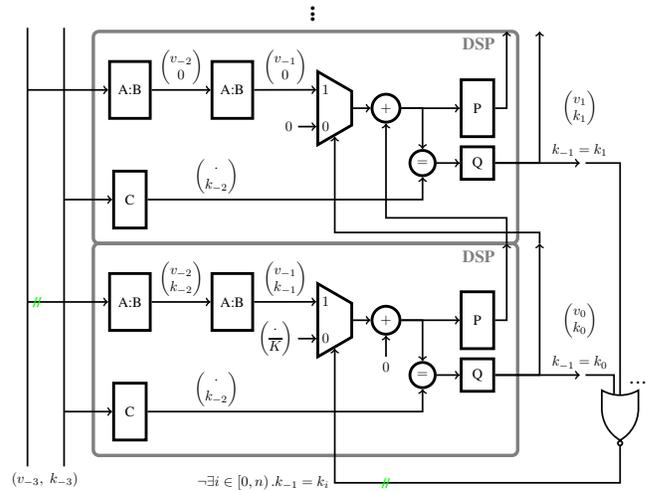
The adjusted mapping to actual DSP slices is illustrated in
\rfig{figDSPMapping}. Note that the indeces used in key-value pairs
designate their age with index zero aligning with the output of the
first DSP slice. The input pipelining within the DSP slice has the consequence
that the two following inputs have already entered the slices. The shorter
input path through the $C$ register is taken by the latest key for its
distributed parallel matching. The match output then controls the merger
of the corresponding increment value in the \emph{next} DSP slice. For the
proper alignment, this value is delayed by another input register on the $A$:$B$
path. Observe the different function of the MUX-controlled input path in
the first DSP slice. It replaced any update that was matched by another
valid queue entry by an empty slot marked by $\bar{K}$.

Again referring to \rfig{figDSPMapping}, note that the main pipeline path is
entirely internal to the chain of DSP slices by using designated cascading
busses for forwarding the output registers $P$. These wide, 48-bit busses are
typically able to carry the whole payload comprising (a) the key extended by a
valid bit and (b) the increment value. For a plain counting statistics, the
value range is bounded by the requirements of merging increments, i.e.
$\log_2(N)$ where $N$ is the depth of the pipeline. Thus, 9 or 10~bits
suffice for a queue matching the memory read latency of a few hundred cycles.
Discounting the extra flag added to identify empty pipeline slots, this
leaves enough space to support a key space of $2^{37}>10^{11}$, i.e. more
than 100~billion.

The challenges, this DSP mapping leaves unsolved are (a) the fanout of the
input pair and (b) the fanin for combining all the individual match results.
As the input is latency-insensitive, duplicated input buffers can be used to
limit the fanout of individual signals so as to maintain the needed clock
speed. Such a trivial pipelining solution is not available for the feedback of
the match signals as it forms a cycle involving the decision of admitting an
active or an empty slot to the pipeline. A growing hierarchy of LUTs
as required for fanin cones of more than six match inputs is prone to
become the critical path of the design.

The overall match reduction can be pipelined by inserting an equivalent
number of register stages both (a) in the feedback path and (b) on the
key-value input path of the first DSP slice as shown by the green markers
in \rfig{figDSPMapping}. The delay added to the feedback path allows to
pipeline the match aggregation as needed. The corresponding delay on the
input path ensures that the invalidation of matched inputs is
applied to the correct key-value pair. Note, however, that the key-value
pairs within the input delay stages are not available for their comparison
with the following candidate. Unmitigated, this implies a functional hazard
if the spacing between identical keys in the input stream is smaller than
the number of input delay stages.

A possible measure to obtain the required spacing of input keys is a distancing
in stages. A first short conflation queue without a pipelined match aggregation
imposes no specific key-spacing requirements on its input. Having a depth of $N$
comparator stages, it, however, guarantees that any key emitted at its output will
be followed by $N$ different keys or empty pipeline slots. Any matching key that
occurred within this window in the input stream will have been merged with its
preceding match. Limiting the depth $N$ of the initial conflation queue, for example,
to the fanin of a device's LUTs ensures a short critical path while also providing
the freedom to pipeline the subsequent conflation stage accordingly. Multiple
stages may be chained to increase the spacing of identical keys step by step. Observe
that the original task of identifying updates with unique keys to initiate and,
finally, complete read-modify-write cycles is the sole responsibility of the
last of these stages.

For a final operational note, observe that the described approach implements
a \emph{forward} conflation of
updates with a common key. This choice guarantees important properties that an
alternative \emph{backward} conflation could not deliver. It guarantees
(a) that updates are not starved by being repeatedly retracted to later ones
with a matching key,
(b) that accumulated conflated increments are bounded and
(c) that the order of unique keys once committed to the conflation queue is maintained.
Thus, the arithmetic datapath can be dimensioned precisely, and the association
with memory replies occurs naturally by sequence.

\section{Experimental Configuration}
We employ the described update conflation to maintain an online occurrence histogram
over the items of a continuous data stream in the off-chip RLDRAM memory \cite{rldram:2015}
of a VCU118 prototyping board \cite{ug1224}. We distinguish $2^{24}>16\cdot10^6$ data
buckets, i.e., keys. The counter maintained for each key comprises a master count and
a shadow count. The shadow counts are targeted during non-disruptive, consistent snapshot
readouts. They are merged into the master count once a snapshot operation completes. The
designation of an event occurrence to either go directly to the master count or temporarily to the
shadow count during snapshot readouts must be performed as long as their temporal order is
intact, i.e., before the conflation queue performs mergers with outstanding increments.
Therefore, two separate 10-bit counts constitute the value part of the conflation datapath.
The total occupation of the 48-bit datapath amounts to $24+1+2\cdot10$~bits for the key, the
valid flag as well as the two count lanes, respectively.

\begin{table}
  \caption{Characteristics of the 0-6-250-Conflation Schedule}
  \label{tabSchedule}
  \[\begin{array}{crrrr}\toprule
  \mbox{Stage} & \mbox{Gap in} & \mbox{Matchers} & \mbox{Gap out}\\
  \mbox{\#$i$} & k_i & n_i & k_i+n_i\\\midrule
  0 &  0 &    6 &   6\\
  1 &  6 &  244 & 250\\\bottomrule
  \end{array}\]
\end{table}
\rtab{tabSchedule} characterizes the implemented conflation schedule. It comprises
two stages implementing the DSP mapping of \rfig{figDSPMapping}:
(a) an initial short conflation with a purely combinatorial match feedback, and
(b) a deep conflation with $244$ parallel matchers. The initial stage pre-processes
the stream of updates to ensure that there is a minimum of $6$ unrelated slots
in between any two identical keys. This enables the introduction of six pipeline stages
into the critical match feedback path of the following essential deep conflation.

Observe that the depths of cascaded conflation stages can increase quickly as
the critical match reduction can leverage pipelined OR trees of logarithmic
height. As the height of this tree grows, the challenging signal crossings out of
and into the DSP slices are also more and more separated for an additional benefit
to timing closure. As a consequence of the logarithmic growth of the tree height,
no practically relevant implementation will demand more than three cascaded
conflation stages.

The implementation of deep conflation stages needs to take measures to break
the chain of DSP slices. On the FPGA device, the DSP slices are arranged in
columns that also facilitate the designated direct cascade datapaths. The height of
these columns is physically bounded. In the case of the super logic regions (SLRs)
constituting the XCVU9P device of the VCU118 board, there are 120~DSP slices chained in
a column. In fact, we break the pipeline into smaller segments, each comprising up to
42~comparators, to prevent the formation of a monolithic placement unit with
an extreme aspect ratio. This way, pipeline segments can be placed on neighboring
DSP columns to form a more compact, logically interleaved structure. To facilitate
feasible routing between these segments, we unbalance the inter-segment match reduction
to enable the displacement of pipelining fronts to the inter-segment links of the
DSP pipeline.

As a reference, we evaluate providing the same functional capability through
general-purpose fabric logic. In this scenario, the state of the conflation
queue is maintained entirely in fabric registers and, thus, available for a
fully unrolled, parallel key matching in fabric LUTs. A competitive clock
speed is attained by implementing an analogously staged conflation scheme
comprised of (a) a stage with a short combinatorial match feedback, and
(b) a deep conflation stage. Thus, we essentially contrast the trade-off
between different FPGA resource types.

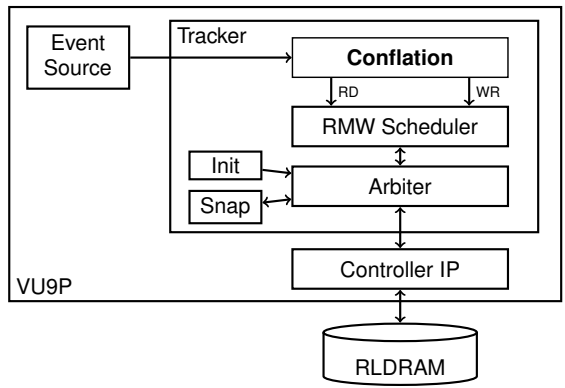
\begin{figure}
    \centerline{\resizebox{.86\linewidth}{!}{% Author: Thomas B. Preußer
\usetikzlibrary{calc}
\usetikzlibrary{shapes.geometric}

\begin{tikzpicture}[scale=.9,line width=1.2,font=\large\sffamily]
\draw (0,-1.8) node[above right] {VU9P}  rectangle (12,4.6);

\node(nsrc)[draw] at(1.5,3.5) {\begin{tabular}{c}Event\\Source\end{tabular}};

\draw (11.5,-.3) rectangle (3.5,4.3) node[below right] {Tracker}; {
  \node(ncfl)[draw,thick, minimum width=120,minimum height=22] at(8.5,3.5) {\textbf{Conflation}};
  \draw[->](nsrc)--(ncfl);

  \node(nsch)[draw, minimum width=120,minimum height=22] at(8.5,2) {RMW Scheduler};
  \draw[->]($(ncfl.south)-(1.5,0)$)--($(nsch.north)-(1.5,0)$) node[midway,right]{\footnotesize RD};
  \draw[->]($(ncfl.south)+(1.5,0)$)--($(nsch.north)+(1.5,0)$) node[midway,right]{\footnotesize WR};

  \node(narb)[draw, minimum width=120,minimum height=22] at(8.5,.7) {Arbiter};
  \draw[<->](nsch)--(narb);

  \node(ninit)[draw, minimum width=40] at(4.7,0|-narb.north) {Init};
  \node(nsnap)[draw, minimum width=40] at(4.7,0|-narb.south) {Snap};
  \draw[->](ninit) -- (narb);
  \draw[<->](nsnap) -- (narb);
}
\node(nctrl)[draw, minimum width=120,minimum height=22] at(8.5,-1.1) {Controller IP};
\draw[<->](narb)--(nctrl);

\node(nmem) at($(nctrl)-(0,2)$) [cylinder,draw,aspect=.17,rotate=90,minimum height=12mm,minimum width=30mm]{\rotatebox{270}{\raisebox{-8pt}[0pt][0pt]{RLDRAM}}};
\draw[<->](nctrl)--(nmem);
\end{tikzpicture}}}
    \caption{Experimental Application Setup}
    \label{figApplication}
\end{figure}
All design evaluations have been conducted in the minimal application context
depicted in \rfig{figApplication}. A trivial pseudo-random \emph{Event Source} is
monitored for occurrence statistics. The event stream is fed to the described
two-stage update \emph{Conflation}, which triggers the start and the completion of
read-modify-write cycles. The \emph{RMW Scheduler} takes the memory transaction
requests into the slower memory user clock domain while also trying to group reads
and writes up to (a) fill the four parallel memory command lanes and (b) to
reduce bus turnaround overhead. It guarantees that reads do not surpass writes
in this grouping process. The \emph{Arbiter} allows to funnel in the administrative
memory access for (a) the counter initialization and (b) the low-priority
snapshot readout. A Xilinx \emph{Controller IP} \cite{pg150} implements the last layer
before going to the off-chip RLDRAM memory device. The memory is operated
at 1066\,MHz. The controller exposes an interface of four parallel command
lanes and four doubled data lanes with a user clock of 133.25\,MHz.
For design synthesis and
implementation, Vivado~2019.2 has been used with default flow settings.

\section{Evaluation}\label{secEvaluation}
\begin{table}
  \caption{Resource Utilization by the 0-6-250 Conflation Schedule}
  \label{tabResources}
  \newcommand{\entry}[1][]{%
   \begin{tikzpicture}[#1]%
    \draw[->] (0,0.8ex) -- (0,0) -- (0.8ex,0);
    \draw (0.65ex,0.15ex) -- (0.8ex,0) -- (0.65ex,-0.15ex);
    \node at(-.1ex,-.1ex) {};
   \end{tikzpicture}%
  }
  \definecolor{shade}{gray}{0.9}
  \resizebox{\linewidth}{!}{\begin{tabular}{lrrrrrr}
  \toprule
  \multicolumn{1}{c}{\textbf{Freq.}}
  &\multicolumn{2}{c}{\textbf{LUTs}}
  &\multicolumn{2}{c}{\textbf{FFs}}
  &\multicolumn{2}{c}{\textbf{DSPs}}\\
  \multicolumn{1}{c}{$[$MHz$]$}
  &Mapped&Ref.&Mapped&Ref.&Mapped&Ref.\\
  \midrule
  350 - Total                           & 13008 & 18562 & 11096 & 21545 & 260 & 3 \\
    \entry       {\footnotesize Tracker}&  1280 &  6837 &  1974 & 12423 & 257 & 0 \\
  \rowcolor{shade}
  ~~\entry\textbf{\footnotesize Confl.} &   561 &  6128 &   982 & 11406 & 257 & 0 \\
    \entry       {\footnotesize MemCtrl}& 11289 & 11291 &  8371 &  8371 &   3 & 3 \\
  \midrule
  360 - Total                           & 13005 & 18553 & 11137 & 21545 & 260 & 3 \\
    \entry       {\footnotesize Tracker}&  1279 &  6832 &  2015 & 12423 & 257 & 0 \\
  \rowcolor{shade}
  ~~\entry\textbf{\footnotesize Confl.} &   562 &  6128 &  1021 & 11406 & 257 & 0 \\
    \entry       {\footnotesize MemCtrl}& 11289 & 11287 &  8371 &  8371 &   3 & 3 \\
  \midrule
  375 - Total                           & 13010 & 18558 & 11218 & 21545 & 260 & 3 \\
    \entry       {\footnotesize Tracker}&  1280 &  6833 &  2096 & 12423 & 257 & 0 \\
  \rowcolor{shade}
  ~~\entry\textbf{\footnotesize Confl.} &   570 &  6130 &  1078 & 11406 & 257 & 0 \\
    \entry       {\footnotesize MemCtrl}& 11294 & 11287 &  8371 &  8371 &   3 & 3 \\
  \midrule
  390* - Total                          & 13010 & 18556 & 11300 & 21545 & 260 & 3 \\
    \entry       {\footnotesize Tracker}&  1281 &  6830 &  2178 & 12423 & 257 & 0 \\
  \rowcolor{shade}
  ~~\entry\textbf{\footnotesize Confl.} &   563 &  6128 &  1160 & 11406 & 257 & 0 \\
    \entry       {\footnotesize MemCtrl}& 11293 & 11294 &  8371 &  8371 &   3 & 3 \\
  \bottomrule
  \multicolumn{7}{l}{%
    *\footnotesize $A$:$B$ cascading of DSP mapping reduced to single re-use.
  }
  \end{tabular}}
\end{table}
The resource utilization obtained for the experimental setup under different
sets of parameters is tabulated in \rtab{tabResources}. It shows the consumption
of the general-purpose fabric resources in terms of LUTs and FFs (registers) as
well as DSP slices for both (a) our proposed conflation mapping and (b) the
reference solely relying on fabric resources. All experiments have been conducted
for four challenging application frequencies from 350\,MHz to 390\,MHz. Each
result is given with its relevant structural context as stated by Vivado's
hierarchical utilization report. While the memory controller dominates
the overall application, the update conflation is, indeed, the most costly
part of the event tracker.

First, observe that both the LUT and the DSP slice utilization are largely
independent from the choice of the target application frequency. The proposed
DSP mapping is a clear trade-off investing DSP slices in exchange for freeing
general-purpose LUTs. It is entirely carried by the implementation of the
conflation queue freeing 5550~LUTs for an investment of 257~DSP slices,
i.e., about 22~LUTs per DSP slice. Note that only 250 of these DSP slices perform
the complete functionality within the conflation queue. The remaining seven
facilitate internal segmentation of the deep second conflation stage and the
final delayed addition at the end of each stage.

A similar resource trade-off is observed for the register utilization. The
investment of 257~DSP slices frees more than 10,000 fabric registers, i.e.
about 39~FFs per DSP slice. Observe that the register pay-off is higher
for lower application frequencies. In contrast to the reference fabric design,
the DSP mapping provokes some register duplication as the target frequency
increases. This can be attributed to the more challenging input routing to the
DSP computation as it produces a greater aspect ratio dictated by the physical
columnar arrangement of DSP slices on the device. In addition to this automatic
inflation of datapaths, going above 375\,MHz also mandated limiting the
re-use of the $A$:$B$ inputs across cascading busses to only two adjacent
DSP slices. Leveraging fewer routing capabilities internal to the DSP slices,
returns more and more routing load to the fabric, thereby, diminishing one the
benefits of moving into the DSP slices. Our results suggest that going beyond
400\,MHz is not practical for the given VU9P device.

The RLD3 memory of the VCU118 board has a peak performance of 1066\,MT/s\footnote{%
    MT/s -- million transactions per second, a transaction being one independent read or write
    of a 72-bit (single device) or a 144-bit (dual device) word
}. This performance is reduced by bank cycle times, bus turnarounds and
refresh interference. Implementing an access schedule alternating between blocks of
16~reads and blocks of 16~writes with a perfect cycling through the 16~available banks
reduces the achievable transaction rate to below 60\% of the nominal peak value.
% For a longer version, we might want to turn this into a graph: block size -> throughput.

%\begin{figure}
%    \centering
%      \begin{tikzpicture}
%      \begin{axis}[
%      width=0.9\columnwidth,
%      height=0.55\columnwidth,
%      xlabel={\#Block Size (Reads and Writes)},
%      ylabel={MT/s},
%      ymin=100,
%      ymax=400,
%      ymajorgrids=true,
%      xmode=log,log basis x={2},
%      log ticks with fixed point,
%      xtick={4, 8, 16, 32, 64, 128, 256, 512},
%      %xticklabel={4,8,16,32,64,128,256,512},
%      xmax=600,
%      %xticklabel style={rotate=40, anchor=north east},
%      xlabel style={at={(0.5,0)}},
%      scaled x ticks = false,
%      ]
%     \addplot+[color=red, line width=1.2pt] coordinates {
%     (4,142.50)
%      (8,194.13)
%      (16,284.95)
%      (32,328.56)
%      (64,341.98)
%      (128,348.87)
%      (256,352.38)
%      (512,354.21)};
%
%      \end{axis}
%      \end{tikzpicture}
%      \caption{RLDRAM User Interface Trasactions vs. \#Block size}
%      \label{fig:rldram_bw}
%\end{figure}
%
A conflating event tracker running at 375\,MHz has a peak
requirement of 750\,MT/s as up to one read and one write operation can be emitted
per cycle. This is a nominal utilization of 70\% of the peak memory bandwidth. It will,
thus, not constitute the limiting bottleneck of the statistics computation. In
terms of operational performance, the fully-associative conflation with
250~comparators and processing events at 375\,MHz amounts to
93.75\,G comparisons per second.

The described conflation approach can be implemented feasibly both on the basis of
the proposed DSP mapping and in the general-purpose fabric. The DSP mapping provides
a valuable resource choice in application contexts that are otherwise dominated by
fabric logic. Leveraging dormant DSP resources removes the pressure imposed by
significant state, comparator logic and pipeline datapath routing from the fabric.
In fact, the resource usage can be finely tuned to complement other demands by
combining both implementation options. A hybrid approach combining a DSP chain with
some parallel state in the fabric would also allow to extend the datapath with a
fine granularity beyond the hardened 48-bit DSP accumulation bus.

Very similar DSP slice architectures as used here are available on virtually all
contemporary devices offered by different vendors on the market. They share the
critical feature of designated cascading datapaths that enable the implementation
of an internal pipeline. These datapaths differ somewhat in bit width. Stratix\,IV
devices by Intel provide 44~bits \cite{stx4hb}, their Stratix\,V devices already
feature 64~bits \cite{wp01131}. However, neither of the Stratix DSP blocks provides an internal
comparator so that the key must be fed from the DSP datapath into the fabric for
the match computation.

Consider as an alternative, the implementation of the update conflation based
on a CAM for the keys leveraging the CAM emulation approach by Jiang
\cite{jiang:2013}. Having no need for wildcard matching, the update delay
can, indeed, be reduced from $O\left(2^w\right)$ to $O(1)$. Using distributed
LUTRAM with an individual address width of $w=5$ results in the consumption of
$\left\lceil\frac{24}{5}\right\rceil\cdot244=1220$~LUTs for the tag memory
equating the deep second conflation stage. $244$~LUTs are needed to complete the
computation of the local key matches into a 244-bit match vector. In order to
support both a key entry and a key deletion within each clock cycle, a second
port must be emulated by memory duplication with an exclusive OR to form the
true output from both LUTRAM sibblings. This results in nearly 3000~LUTs for
the key storage and local matching. In contrast to the conflation pipeline,
there is no propagation of key-value pairs but the storage is operated as a
ring buffer. As the CAM organization supports fast key matching but not
retrieval, an additional block RAM buffers the keys for their fast unqueuing
and deletion. Another block RAM stores the associated increment values. These
memories must be operated in dual-port mode so as to allow the parallel queuing
and unqueuing or the conflation with an outstanding buffered increment. Adopting
the initial preparative conflation stage of our proposal would enable the
attractive pipelining of the needed memory updates. In terms of resources,
this approach would invest two block RAMs for freeing about 3000~LUTs.
% If we get data collected: Characterize read latency histogram.

In summary, our results show that DSP slices offer a capable resource option
for implementing the considered non-blocking hazard prevention. They provide
a computational base that is able to perform vastly parallel key matching and
on-the-fly mergers of matching updates across the off-chip memory read latency.
The described DSP mapping enables designers to designate specifically the least
constrained resource type of their application to the challenging maintenance
of off-chip statistics over high-bandwidth data.

\section{Conclusions}\label{secConclusion}
This paper has described the FPGA implementation of a fully-associative
key-value mapping with an additive update conflation that serves
the hazard avoidance of a high-bandwidth
off-chip statistics collection. It
delivers the massively parallel matching of new entries against all pending keys
of a high-latency read-modify-write cycle so as to match the memory bandwidth.
The proposed conflation scheme has been
shown to map well to the DSP slices of contemporary Xilinx devices.
Concrete timing and utilization results have been reported, interpreted and
evaluated for a conflation schedule suitable for collecting event occurrence
statistics in the RLDRAM memories of the VCU118 prototyping board.
They demonstrate that the recourse to DSP slices offers a very capable
resource option freeing valuable general-purpose fabric resources for
implementing a challenging non-blocking associative update mapping for
the hazard prevention in the statistics collection in off-chip memories
with a significant read latency.

%\newpage
\IEEEtriggeratref{13}
\bibliographystyle{IEEEtran}
\bibliography{dsp_assoc}

\end{document}